\documentclass[journal]{vgtc}   
\ifpdf%                                % if we use pdflatex
  \pdfoutput=1\relax                   % create PDFs from pdfLaTeX
  \usepackage{graphicx}                % allow us to embed graphics files
  \DeclareGraphicsExtensions{.pdf,.png,.jpg,.jpeg} % for pdflatex we expect .pdf, .png, or .jpg files
\else%                                 % else we use pure latex
  \usepackage{graphicx}                % allow us to embed graphics files
  \DeclareGraphicsExtensions{.eps}     % for pure latex we expect eps files
\fi%

\graphicspath{{figures/}{pictures/}{images/}{./}} % where to search for the images

\usepackage{microtype}                 % use micro-typography (slightly more compact, better to read)
\PassOptionsToPackage{warn}{textcomp}  % to address font issues with \textrightarrow
\usepackage{textcomp}                  % use better special symbols
\usepackage{mathptmx}                  % use matching math font
\usepackage{times}                     % we use Times as the main font
\usepackage{cite}                      % needed to automatically sort the references
\usepackage{tabu}                      % only used for the table example
\usepackage{booktabs}                  % only used for the table example
\usepackage{balance}       % to better equalize the last page
\usepackage{graphics}      % for EPS, load graphicx instead 
\usepackage[T1]{fontenc}   % for umlauts and other diaeresis
\usepackage{txfonts}
\usepackage{mathptmx}
\usepackage[pdflang={en-US},pdftex]{hyperref}
\usepackage{color}
\usepackage{xcolor}
\usepackage{booktabs}
\usepackage{textcomp}

\PassOptionsToPackage{hyphens}{url}\usepackage{hyperref}

\usepackage{microtype}        % Improved Tracking and Kerning
\usepackage{ccicons}          % Cite your images correctly!
\usepackage{todonotes}

\usepackage{soul}
\usepackage{balance}
\usepackage{enumitem}

\usepackage{multicol}
\usepackage{float}
\usepackage{color, colortbl}
\usepackage{csquotes} %Elsie added this

\setlist{nosep}

\definecolor{lgray}{gray}{0.9}

\usepackage{changes}

\newcommand{\finalstate}[1][]{%
  \renewcommand{\added}[2][]{##2}
  \renewcommand{\deleted}[2][]{}%
  \renewcommand{\replaced}[3][]{##2}}%

\onlineid{0}
\vgtccategory{Research}
\vgtcpapertype{theoretical}
\title{Affective Learning Objectives for Communicative Visualizations}
\author{Elsie Lee-Robbins and Eytan Adar}
\authorfooter{
\item
  Elsie Lee-Robbins and Eytan Adar are with the University of Michigan. E-mail: {\{elsielee, eadar\}@umich.edu.}
}

\shortauthortitle{Lee-Robbins and Adar: Affective Learning Objectives for Communicative Visualizations}
\abstract{When designing communicative visualizations, we often focus on goals that seek to convey patterns, relations, or comparisons (cognitive learning objectives). We pay less attention to affective intents--those that seek to influence or leverage the audience's opinions, attitudes, or values in some way. Affective objectives may range in outcomes from making the viewer care about the subject, strengthening a stance on an opinion, or leading them to take further action. Because such goals are often considered a violation of perceived `neutrality' or are `political,' designers may resist or be unable to describe these intents, let alone formalize them as learning objectives. While there are notable exceptions--such as advocacy visualizations or persuasive cartography--we find that visualization designers rarely acknowledge or formalize affective objectives. Through interviews with visualization designers, we expand on prior work on using learning objectives as a framework for describing and assessing communicative intent. Specifically, we extend and revise the framework to include a set of affective learning objectives. This structured taxonomy can help designers identify and declare their goals and compare and assess designs in a more principled way. Additionally, the taxonomy can enable external critique and analysis of visualizations. We illustrate the use of the taxonomy with a critical analysis of an affective visualization.  
}

\keywords{Affective visualization, communicative visualization, learning objectives.}

\vgtcinsertpkg

\begin{document}
\finalstate


\maketitle
\section{Introduction}

% Cognitive vs affective goals
Data visualization designers often emphasize their goal of conveying facts, insights, comparisons, and patterns to their audience through communicative visualizations. Goals of this type are commonly viewed as `cognitive objectives' (e.g., recall that group X's unemployment is greater than group Y's). By modeling the designer as `teacher' and viewer as `student' it is possible to state intents as learning objectives~\cite{Adar_Lee_2020}. Using a learning objectives framework, a designer can explicitly state their objective (e.g., ``the viewer will analyze the impact of different policy `bundles’ on global temperature'') and assess whether a visualization successfully supports this outcome. Most attention in the data visualization field---both from researchers and practitioners---focuses on the cognitive domain. However, data visualization practitioners also have goals that go beyond the cognitive domain---they want their audience to have a \textit{reaction} or a \textit{response} to their visualization. For example, a designer may want their viewer to consider that Obamacare is a bad system if they show a chaotic diagram~\cite{klein} (or a good one if they show an organized one~\cite{palmer}). The cognitive intent becomes limited in these visualizations; the designer doesn't need the viewer to remember how Obamacare works or critique its particular features. Rather, the designer may want their audience to agree with an appraisal (e.g., Obamacare is bad), accept an attitude (e.g., hospitals should be for-profit), or believe in a value (e.g., small government). These goals are \textit{affective} intents, and cognitive learning objectives do not cover them.

% Affective visualizations in advocacy or social justice
Affective intents are most obvious in data visualizations that are created for advocacy reasons. With these, designers are clearly trying to raise awareness or have a call to action. Advocates for a cause are not trying to hide the fact that they are taking these positions and that they want you to care about their cause too. 
% Periscopic example
The visualization ``U.S. Gun Deaths'' created by Periscopic is a very clear example of an affective visualization \cite{Periscopic_2013}. Among other features, the visualization animates a tally of `stolen years' due to gun violence. The cognitive aspect of the visualization---how many people were killed---is important, but not the main takeaway of the visualization. Periscopic co-founder Kim Rees reflects, ``We need people to react. We need people to sort of get riled up about things, get excited about things, and want to make change in the world'' (\cite{Schwabish_Rees_Mushon_2016}, 5:30). At a minimum, the goal of ``U.S. Gun Deaths'' is to create interest in the topic. Ideally, the visualization will evoke empathy within the audience for the victims. The affective response, not the cognitive component, is the main intent of the visualization.

However, even in domains other than advocacy and social justice, designers have affective goals. They want the viewer to care about the topic, strengthen an attitude, or take further action. A visualization of a family tree might have you consider your own family ties. A visualization about sleep patterns might lead you to value sleep. A visualization about blood donations might inspire you to donate blood. Unfortunately, because of the lack of attention in this area, designers may not even realize or consider that they have affective goals. Given the focus on `neutrality' (or related concepts), persuasive data visualizations are often controversial, which might make designers try to hide the fact that they have affective goals. Even if a designer acknowledges that they have affective goals, they may not have the vocabulary or framework to articulate their goals. 

% In this paper -- summarize study
In this paper, we extend our previous research on cognitive learning objectives to the affective domain, adapting the affective taxonomy from the education realm to work for data visualizations. We conceptualize \textit{affective intents} as goals regarding an audience's \textit{reaction} or \textit{response} to appraisals, attitudes, or values. We begin by adapting the (so-called) Bloom's affective taxonomy\cite{Krathwohl_Bloom_Masia} to provide a framework for communicative data visualization intent. We conducted an interview study to learn more about data visualization practitioners' affective goals, how they conceptualize their intent, and how they could use learning objectives. Based on this study, we revised the affective taxonomy to better align with visualization intents. A language for describing affective intents will not only benefit designers, but will also enable new kinds of critique. We also draw a distinction between affective rhetorical techniques (i.e., `pathos') in contrast to affective intents. Specifically, we contribute: an affective learning objectives taxonomy for data visualization, a qualitative analysis of an interview study of 12 designers, and a critical analysis of an affective visualization with associated learning objectives to illustrate the taxonomy's use. \added{Data visualization designers will be able to apply this framework to their own work to consider their affective intents.}

\section{Background}

% intro
Data visualizations are not neutral, even though designers and viewers might want them to be. 
To create a data visualization, designers must make choices that will shape how the audience will interpret the data. Intentionally or unintentionally, designers have biases, backgrounds, and personal opinions and preferences that will influence their design decisions. 
Strategically, designers employ logical (logos), emotional (pathos), and credibility (ethos) elements that are designed to evoke emotion in the audience (e.g., visual imagery). In many cases, these pathos-based techniques, such as humanizing data, coincide with trying to achieve an affective goal. Though our ultimate goal is to create a language for describing affective \textit{intents}, we begin by reviewing affective \textit{persuasion strategies}. 

Note also that we adopt the language of designer and viewer in this work (per~\cite{Adar_Lee_2020}). The designer here does not necessarily mean the person who created it but may indicate an editor who has chosen it. We also broadly use the idea of a visualization as not only the image but also the context (text, captions, etc.) with which it is associated.

\subsection{Data Visualizations are Rhetorical Objects}

% Rhetorics
Communicative visualizations are rhetorical devices---they are a way to convince an audience of an idea. Narrative visualizations, a form of communicative visualizations, ``tell a story'' and use rhetorical techniques to make an argument \cite{Hullman_Diakopoulos_2011}. Enrico Bertini argues, ``We do not visualize \textit{data}, we visualize \textit{ideas} based on data'' \cite{Bertini_2022} (emphasis in original). With visualizations, the most apparent argument is a logical one---the data is a fact that supports a claim. However, data visualizations are not solely logical arguments. In fact, visualizations can also have an appeal to emotion \cite{Campbell_2018}. Both of these modes work together to persuade the audience to believe the data. 

% Persuasion
Data visualizations can be effective at persuasion \cite{Pandey_Manivannan_Nov_Satterthwaite_Bertini_2014}, but are not always successful \cite{Kong_Liu_Karahalios_2018}. Cartographic visualizations, in particular, have a history of being used for propaganda, political pressure, and persuasion \cite{Tyner_1982}. Designers will choose how a three(+)-dimensional reality will be distorted through two-dimensional maps. 
% Advocacy
Because of their potential to influence people, visualizations are used in advocacy \cite{TactileTechnologyCollective_2013}. Historically, visualizations have be utilized to bring attention to, and provoke action for, many issues. Examples include \deleted{John Snow's cholera map, }Florence Nightingale's military medicine and Charles Minard's Napoleon's march on Russia~\cite{Emerson_Satterthwaite_Pandey_2018}. Paired with a call to action, visualizations can galvanize people to act~\cite{Lambert}. These examples reflect the subtle interaction between rhetorical tools to persuade and the ultimate intent of the visualization (see Figure~\ref{fig:intentVStech}).

\subsection{Data Visualizations are Not Neutral}

% Data visualization is biased
A key motivation for our attempt to characterize affective intents is to challenge the idea the neutrality is desirable, or even possible, in visualization. Data is biased throughout the entire process, from collection, to cleaning, to encoding and communicating with an audience \cite{DIgnazio_Klein_2020, Correll_2019, Thudt_Perin_Willett_Carpendale}. The designer of a visualization has to make editorial choices that will influence how a viewer perceives the message of the graph \cite{Carusi_Hoel_Webmoor_Woolgar_2014}. There are many choices that a designer makes that could affect interpretation, including what variables to encode, what colors to use \cite{Bartram_Patra_Stone_2017}, whether to highlight and annotate, whether to add visual embellishments \cite{Bateman_Mandryk_Gutwin_Genest_McDine_Brooks_2010}, what context to compare the data to, how to frame the title of the visualization \cite{Kong_Liu_Karahalios_2018, Kong_Liu_Karahalios_2019}, and what type of chart to use \cite{Zacks_Tversky_1999}. Whether or not the designer is aware of it, their intents, prior experience, and own personal preferences influence their design choices of the visualization.

% Illusion of objectivity
Even though there is subjectivity in data visualization, there can be an illusion of objectivity for both designers and the audience. Designers can intentionally (or unintentionally) use conventions, such as minimalist aesthetics or the inclusion of data sources, to make data visualizations seem objective, factual, and transparent \cite{Kennedy_Hill_Aiello_Allen_2016}. This perception of neutrality could also result from an association with scientific methods, which are perceived as objective \cite{Cairo_2013}. Designing a visualization as if the audience can see everything from a neutral perspective is what Donna Haraway calls ``The God Trick'' \cite{Haraway_1988}. 
As opposed to text, which has a more apparent narrator perspective, ``images and statistics do not seem to imply an inherent point-of-view'' \cite{Rock_1992}. Thus, data visualizations are often incorrectly perceived by the audience as neutral and objective \cite{Kong_Liu_Karahalios_2018, Kong_Liu_Karahalios_2019, Pirani_Ricker_Kraak_2020, vanKoningsbruggen_Hornecker_2021}. 
Although the audience might not realize it, data visualizations can be created for malicious intents \cite{Correll_Heer_2017}, deceive or mislead \cite{Cairo_2019}, be intentionally slanted \cite{Kong_Liu_Karahalios_2018}, and be used to support both sides of an argument \cite{Lee_Yang_Inchoco_Jones_Satyanarayan_2021}. For visualization designers who have \textit{acknowledged} affective intents, an appropriate language for describing these can support better design and critique.

\subsection{Neutrality-Adjacent Goals}

% People have goals of objectivity
Although we have argued that data visualizations are not neutral, some practitioners will still contend that their goal is to represent data in an unbiased way as possible. Some may view their job as to inform their audience in a neutral way, without any bias or persuasion. For these \textit{unacknowledged} affective goals, we propose that an affective learning objective language will better enable self-critique and analysis (even an ability to express what the designer doesn't want).

% Due objectivity in journalism 
Members of the journalism field are strong proponents of what we term neutrality-adjacent goals. These come from the underlying idea of representing all sides of a situation fairly and the pursuit of truth. Impartiality is a common goal for journalism \cite{NPR, NYT}. An objective for an election visualization could be \textit{giving equal representation to all parties}. This could mean choosing colors that are of equal salience, or to frame the title to label the visualization's topic rather than pointing out one candidate or the other. 
A modified guideline, \textit{due objectivity} \cite{BBC, Ofcom_2021}, also takes into consideration context, harm, and potential outcomes, giving each side the platform that they deserve. This may result in only presenting one side. In the case of climate change, for example, some news organizations no longer give climate-deniers a platform to present their opinions. 

% Neutral affect
In some highly polarized political debates, a designer may want to strip all emotion out of the presentation with an objective to present the data with \textit{neutral affect}. Sandra Steingraber, an author and biologist, noted that, ``there are times, actually, when communities are so divided emotionally over an issue that I actually feel my task is simply to speak plainly about science $\ldots$ In those cases I actually keep affect out''~\cite{Slovic_Slovic_2015}. This goal is not saying that the visualization itself is neutral, rather that the audience will not have an emotional reaction to the visualization. 
In a similar vein, a designer could have a goal of appearing non-political. For a politicized topic, they may try to remove indicators that bring awareness to the political nature of the topic. 

% Objectivity as a process
A designer's goal could be to represent the data as accurately and truthfully as possible, intending an \textit{objectivity of method} \cite{Kovach_Rosenstiel_2001}. This goal acknowledges that designers are subjective beings, but that a scientific method will bring a level of objectivity to the content that a data visualization designer communicates (or a journalist, in the original context of \cite{Kovach_Rosenstiel_2001}). In this way, objectivity is not a state, but rather a process of collecting the data in the most accurate and faithful way possible, representing it to avoid misperceptions of the data. 

Neutrality (and neutrality-adjacent) goals connect to affective intents--we are often trying to convince someone to believe something. We contend that many of these neutrality-adjacent goals (many of which lead to non-neutral effects) would benefit from a clear language to describe affective intents.

\subsection{Emotion in Data Visualizations}

% Introduction
In this work, we focus on affective intents, defined as goals relating to a viewer's appraisals, attitudes, values, or value system. The term \textit{affective} in research literature could alternatively refer to mood, feelings, and emotions. Emotion is commonly found in affective data visualizations, but is not necessary nor sufficient on its own. Emotions may play a role in how designers achieve their goal, but our conceptualization of affective intents is on the resulting change in appraisals, attitudes or values. While we discuss emotions, we do not include them in our taxonomy.

\begin{figure}[htb]
  \centering
\includegraphics[width=\linewidth]{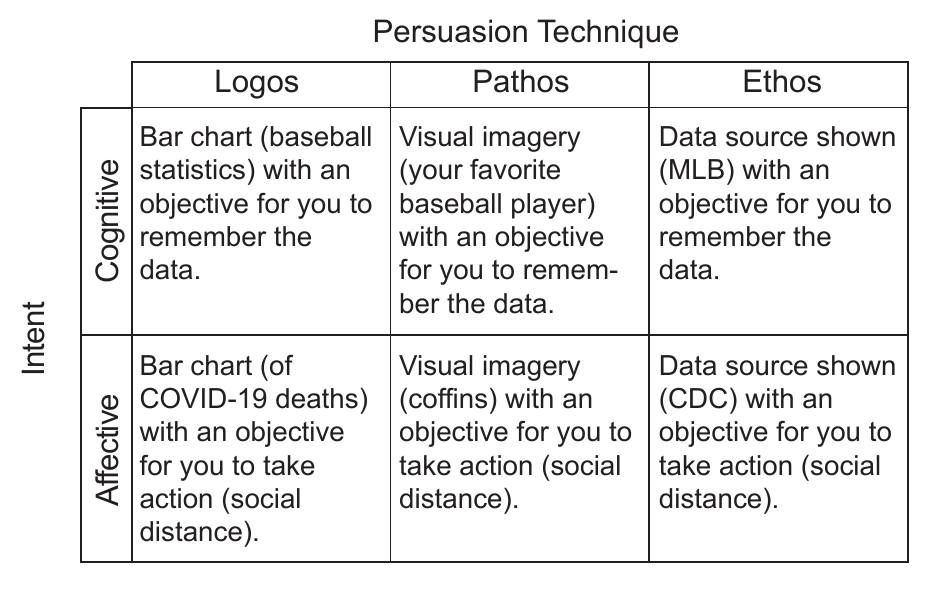}
\vspace{-.5cm}
  \caption{Examples of persuasion techniques (logos, pathos, ethos) used to achieve intents (cognitive, affective).}
\label{fig:intentVStech}
\end{figure}

% Should / can data visualizations evoke emotion
Empathy in particular has been a controversial target in discussions of whether data visualizations \textit{should} convey or evoke emotion \cite{Zer-Aviv_2015, Cairo_2013, Lupi_2018, DIgnazio_Klein_2020}. 
Nonetheless, designers often use pathos techniques in data visualization as a way to achieve their affective intent. We view evoking emotion as a technique, whereas the resulting attitude change is the goal. We illustrate the ways that cognitive and affective intents differ from logos, pathos, and ethos strategies in Figure \ref{fig:intentVStech}. While there has not been as much attention focusing on affective intents, there is more research about pathos techniques. We briefly summarize this literature, but note that a comprehensive review of all techniques is out of scope for this paper. Additionally, we review ways that emotions and attitudes have been measured in the context of data visualization. 

\subsubsection{Pathos Techniques in Data Visualizations}

% anthropographics
A subset of affective visualizations can be classified as \textit{anthropographics}, which are ``visualizations that represent data about people in a way that is intended to promote prosocial feelings (e.g., compassion or empathy) or prosocial behavior (e.g., donating or helping)'' \cite{Morais_Jansen_Andrade_Dragicevic_2020}. 
Some design techniques in anthropographics to \textit{humanize} the data are to 
emphasize specific individuals \cite{Harris_2015, Rost_2017, Bui, Morais_Jansen_Andrade_Dragicevic_2020}, 
make marks more realistic \cite{Harris_2015, Rost_2017, Boy_Pandey_Emerson_2017, Bui, Morais_Jansen_Andrade_Dragicevic_2020}, 
show each individual \cite{Rost_2017, Boy_Pandey_Emerson_2017, Morais_Jansen_Andrade_Dragicevic_2020},
or only focus on a subset of the data \cite{Morais_Jansen_Andrade_Dragicevic_2020}. 
The effectiveness of anthropographics has been researched, with results suggesting at most, a small effect of these design techniques increasing monetary donations \cite{Boy_Pandey_Emerson_2017, Morais_Jansen_Andrade_Dragicevic_2021}. 

% Affective engagement
Pathos techniques can also focus on evoking an emotional response to \textit{engage} the audience \cite{Campbell_2018}. These design elements can grab a viewer's attention, evoke emotion, and create a more engaging experience.
Some techniques that can be used for engagement are 
color or mood \cite{Kostelnick_2016, Bartram_Patra_Stone_2017, Rost_2017, Lan_Shi_Zhang_Cao_2021},
visual imagery embellishments or art \cite{Bateman_Mandryk_Gutwin_Genest_McDine_Brooks_2010, Montanez_2015, Kostelnick_2016, Lan_Shi_Zhang_Cao_2021}, narratives or stories \cite{Kostelnick_2016, Alamalhodaei_Alberda_Feigenbaum_2020, Lan_Shi_Zhang_Cao_2021}, 
visceralization or sensory experiences (e.g., audio, physicalizations) \cite{Kostelnick_2016, DIgnazio_Klein_2020, Elli_Bradley_Collins_Hinrichs_Hills_Kelsky_2020},
immersion or virtual reality \cite{Ivanov_Danyluk_Jacob_Willett_2019, Romat_HenryRiche_Hurter_Drucker_Amini_Hinckley_2020}, 
audience participation or interactivity \cite{Kostelnick_2016, Perovich_Wylie_Bongiovanni_2020, Alamalhodaei_Alberda_Feigenbaum_2020},
animation \cite{Kostelnick_2016, Lan_Shi_Wu_Jiao_Cao_2021}, 
novelty \cite{Kostelnick_2016, Lan_Shi_Zhang_Cao_2021},
and personalization or proximity \cite{Kostelnick_2016, Campbell_Offenhuber_2019}.

\subsubsection{Measuring Emotions, Attitudes, and Behaviors}

Emotion is clearly important for achieving affective intents and various studies have sought to measure the interaction between emotions and visualizations. Study participants in these experiments self-report how strongly they felt emotions \cite{Boy_Pandey_Emerson_2017, Campbell_Offenhuber_2019, Morais_DandaraSousa_Andrade_2020, Pirani_Ricker_Kraak_2020},
label the visualization with affective words \cite{Lan_Shi_Zhang_Cao_2021}, or 
report their awareness of affective aspects \cite{Lan_Shi_Zhang_Cao_2021}.
%Attitudes
Alternative measurements include asking participants 
to report their attitudes on a Likert scale (strongly agree to strongly disagree) \cite{Campbell_Offenhuber_2019},
how likely they are to donate money \cite{Boy_Pandey_Emerson_2017},
or to allocate hypothetical donations \cite{Boy_Pandey_Emerson_2017, Morais_DandaraSousa_Andrade_2020}.
Additionally, researchers can use standard measurements that are validated for measuring specific attitudes (e.g., attitudes toward immigration \cite{Liem_Perin_Wood_2020}).
% Qualitative
Researchers can also use qualitative methods, such as focus groups \cite{Kennedy_Hill_2018}, interviews \cite{Peck_Ayuso_El-Etr_2019, vanKoningsbruggen_Hornecker_2021}, or open-ended responses \cite{Pirani_Ricker_Kraak_2020} to gain an understanding of emotional reactions and attitudes. 

\subsubsection{Cognition and Decision Making}

% emotion bad
Emotion is historically perceived to be at odds with rationality. Paul Bloom argues that relying on empathy is a bad basis for decision making \cite{Bloom_2016}. In any case, emotions influence decision making and mediate behavioral responses \cite{Lowenstein_Weber_Hsee_Welch_2001}. Emotion (from pathos techniques) can help achieve \textit{cognitive} intents (see Figure \ref{fig:intentVStech} for examples).
% Pathos techniques for Cognitive Intents
Visual imagery embellishments, also known as `chart junk', have been found to increase memorability of data visualizations \cite{Bateman_Mandryk_Gutwin_Genest_McDine_Brooks_2010, Borkin_Vo_Bylinskii_Isola_Sunkavalli_Oliva_Pfister_2013}.

\subsection{Learning Objectives}
The idea of utilizing \textit{cognitive} learning objectives to describe designer intents was introduced in earlier work~\cite{Adar_Lee_2020}. This taxonomy enables designers to structure their intents in simple \textbf{The viewer will \textit{[verb] [noun]}} statements. Both the verbs and nouns are hierarchically organized\footnote{see \url{visualobjectives.net}}. For example, verbs range from remember, to understand, to apply, and so on. Nouns range from factual to meta-cognitive. Learning objectives can help visualization designers choose more effective designs \cite{LeeRobbins_He_Adar_2022}.
% cognitive learning objectives in data viz
Learning objectives have been used in data visualization to evaluate user engagement \cite{Mahyar_Kim_Kwon}, visual analytic processes \cite{Sperrle_Jeitler_Bernard_Keim_El-Assady_2021}, specific chart types (e.g., pictographs \cite{Burns_Xiong_Franconeri_Cairo_Mahyar_2021}, 3D visualizations \cite{Saenz_Baigelenov_Hung_Parsons}), and multiple levels of understanding \cite{Burns_Xiong_Franconeri_Cairo_Mahyar_2020}. 

% Affective objectives
However, the cognitive learning objectives framework is insufficient for affective communicative intents. For example, visualization designers can also intend for their viewer react or respond to an appraisal, attitude, or value. Designers can also have goals such as prompting action or starting a conversation about the data \cite{Knaflic_2015}. Thus, we extend this prior literature to include affective learning objectives for data visualization communicative intents.

\section{An Affective Taxonomy for Education}

% Cognitive objectives
The authoritative handbooks on learning objectives are widely accepted to be Bloom's taxonomy \cite{bloom1956taxonomy,Krathwohl_Bloom_Masia}. In Bloom's broader framework, learning is separated into three domains: cognitive (i.e., acquiring and using knowledge), affective (i.e., attitudes and appreciation), and psychomotor (i.e., physical movements of the body). The cognitive domain is the most straightforward and common---the teacher wants the student to be able to know or do something after the class. Because it is the most used, the cognitive taxonomy has been evolved and refined significantly~\cite{anderson2001taxonomy}. Thus, it is easier to adapt this taxonomy to model cognitive intents in communicative visualizations~\cite{Adar_Lee_2020, LeeRobbins_He_Adar_2022}.

Less frequently considered, yet still important, is the affective domain~\cite{hauenstein1998conceptual, Krathwohl_Bloom_Masia}. Instructors have goals for their students to have an appreciation for the course subject, be motivated or interested in learning, or develop a favorable attitude towards the subject. These are different than cognitive learning objectives, as they focus on moods, feelings, or attitudes. Affective learning objectives have received less attention for several reasons. First, affective learning is hard to measure. A multiple-choice question to gauge interest or appreciation would likely be easy for the student to lie about. Second, affective learning takes a long time. Developing a positive association for a subject could take years, or a whole lifetime of learning, making it again difficult to measure. Finally, affective goals can have a negative stigma; in education, teachers could be accused of indoctrination. This combination of factors leads to a less refined taxonomy, but a viable starting point. In our past work~\cite{Adar_Lee_2020}, we made a brief proposal for this model$^1$.

\begin{figure}[htbp!]
  \centering
\includegraphics[width=\linewidth]{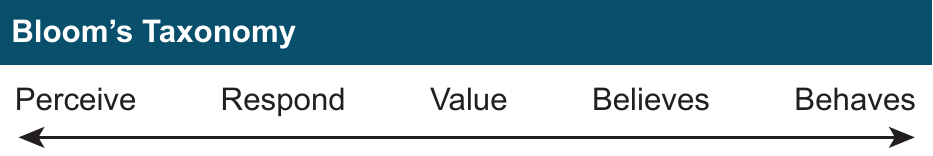}
\vspace{-.5cm}
  \caption{Bloom's Affective Taxonomy (adapted from ~\protect\cite{hauenstein1998conceptual})}
\label{fig:bloom}
\end{figure}

\subsection{Affective Verbs}
% Verbs
In the original taxonomy, the authors conceptualized the main dimension of learning along the concept of \textit{internalization} \cite{Krathwohl_Bloom_Masia}. At the very minimal end of the goals, the viewer will \textit{receive} (or \textit{perceive} in~\cite{hauenstein1998conceptual}) the content that the visualization is presenting (e.g., the viewer will perceive the harm of plastic pollution). At this point, the act of \textit{attending to} is the bare minimum, with interest being a common lower-level goal. Perceiving is a necessary prerequisite for deeper levels of engagement. As the viewer is attending to the visualization, the next level would be for them to \textit{respond}. A response could be a willingness to comply or starting to experience enjoyment (e.g., the viewer will enjoy recycling). A deeper level of internalization is for the viewer to \textit{value}. An example could be agreeing with an opinion or accepting a value (e.g., the viewer will agree that reducing one's carbon footprint is essential). The next step is \textit{organization} (this level is \textit{believes} in~\cite{hauenstein1998conceptual}), which is to organize values into a value system, rank values and evaluate which ones are more important (e.g., the viewer will value sustainability over convenience). At the maximal level of affective learning, there is \textit{characterization} (called \textit{behaves} in~\cite{hauenstein1998conceptual}), where the viewer will fully internalize the values. They will behave in agreement with these values and see these values as part of their identity (e.g., the viewer will identify as an advocate for sustainability). We have adopted this main dimension for the verbs of this taxonomy. 

\subsection{Affective Nouns}
The cognitive taxonomy was extended to explicitly account for a hierarchy of cognition-related nouns (e.g., facts, meta-knowledge, etc.~\cite{anderson2001taxonomy}). However, neither the original affective taxonomy nor any extensions attempt to categorize the learning objectives by their ``noun types.'' As an initial starting point, we conceptualized the noun dimension as consisting of opinion, attitude, value, and value orientation. \textit{Opinions} are weakly held beliefs, which may or may not be backed up with evidence (e.g., there is too much plastic pollution in the ocean). Opinions are the easiest to change, and may be the focus of most affective visualizations. With \textit{attitude} we find an evaluative statement of an object that has an affective component (e.g., recycling is good). Although there is no universally held description of an attitude, we use Eagly and Chaiken's definition, where an attitude is ``a psychological tendency that is expressed by evaluating a particular entity with some degree of favor or disfavor''\cite{eagly1993psychology}. \textit{Values} represent a broader notion which guides a more significant set of actions and is applicable across many situations (e.g., sustainability). Bergman describes values as ``relatively stable beliefs $\ldots$ major life goals and general modes of conduct''~\cite{Bergman_1998}. Values enable decision making around what is `good' or `bad'. \textit{Value system} is the final category. A particular value system (e.g., the Green Party's 10 Key Values~\cite{greenparty}) suggests a larger set of held values that represent a group of people or a community. A value system is described as ``the specific architecture, character, and driving force of a group'' \cite{Bergman_1998}. As with cognitive objectives~\cite{Adar_Lee_2020}, this taxonomy is used to construct statements of the form: \textbf{The viewer will \textit{[verb]} \textit{[noun]}}. Note that while the verbs can be used directly (e.g., `value'), the nouns are replaced with specific opinions, attitudes, values or value systems (e.g., `Obamacare is beneficial to the American public' or `progressive values').

\section{Interview Study}

% intro / purpose / summary / overview
To understand if our proposed taxonomy fits actual designer intents, we initiated a semi-structured interview study. The interview was comprised of three main parts: background on the designer and their visualization, our explanation of learning objectives, and the designer creating their own learning objectives for their visualization.

\subsection{Participants}
We recruited participants from the Data Visualization Society's Slack workspace, from the channels \#share-showcase, \#share-critique, and \#topic-data-art. From these channels, we reached out to people who had posted their own data visualization. Additionally, we only reached out to those that potentially had an affective intent, as indicated by the description, topic, or design of the visualization. Because there are far fewer visibly affective communicative visualizations relative to obviously cognitive ones, we also reached out to an additional 3 people on Twitter who posted visualization examples. As an opening introduction, we direct messaged 51 people to ask what their goals were for the visualization. We received 34 responses to our initial message. We used their response to assess again whether there was an affective intent. From there, we invited 17 people to participate in an interview, of which 12 accepted. 

% participant backgrounds
Most of our participants create visualizations as part of their full time jobs. Our participants ranged from 2 years of experience as a hobby to more than 10 years of experience as their main career. Out of the 12 participants, 6 were journalists, 4 had data science backgrounds, 3 had science backgrounds, 3 had graphic design backgrounds, 8 had another type of background (participants could have more than 1 type of background). Half \deleted{(6) }of the participants noted that they had formal education in data visualization, meaning that they mentioned a masters degree, bootcamp, or some other structured learning specifically about data visualization. \added{Since the data visualization society is a global organization, we had participants from New Zealand, India, Finland, The Netherlands, Spain, Singapore, and the United States. }

\subsection{Methods}

% interview summary
The interviews were 60 minutes long, and were conducted over Zoom \added{by the first author}. After gaining consent for participation, participants were presented with \added{a slide desk to follow along throughout the interview}. First, the participant was asked to describe their background and experience in data visualization and their goals for the specific visualization they had posted. Then, we gave a brief explanation of cognitive learning objectives and affective learning objectives. We used an example visualization to illustrate what learning objectives are and how they could be used in data visualization. Finally, the participant was asked to create cognitive learning objectives and affective learning objectives for their own visualization. We asked follow up questions to better understand their experience creating these learning objectives. Participants were offered a \$30 gift card for their time. The slide deck and the set of starting questions we asked are included in Supplementary Materials. 

% modified explanation
This semi-structured interview design is modified from our previous interview study on cognitive objectives~\cite{Adar_Lee_2020}. In that work, we reached out to people with learning objectives that \textit{we} had created for their visualization and had them evaluate if those actually fit their intent. Because we found that some designers are resistant to the idea that they have affective intents, we used a more general initial question this time: \textit{``In as much detail as possible, what were your goals for this visualization? In other words, what effect on your audience were you hoping to achieve with your visualization?''} We made this change because we were interested in how designers would frame their intent in their own words. In contrast to our past study design where participants selected and created objectives at the survey stage, we asked participants to create learning objectives during the interview. We added this section to get a better understanding of how designers would actually create learning objectives. Our expectation was that this would reveal what questions participants have about learning objectives, how difficult it is to create them, and if they would use learning objectives in the future. 

% qualitative analysis
The interviews were recorded and transcribed. \added{We used established qualitative approaches for data analysis\cite{diehl2022characterizing}}. The initial code book was pre-populated with codes from our previous study on cognitive learning objectives \cite{Adar_Lee_2020}. We only used codes that applied to 2 or more participants from the previous study, starting us with 49 codes. Additional codes were added using open coding \added{by the first author} during close readings of the interview transcripts. After all of the interviews were coded, the first author revised the code book, combining similar codes and removing codes that were either too broad or too specific. The interviews were recoded again according to the revised code book. In total, the final code book contains 172 codes. 

% thematic analysis
We conducted a thematic analysis using inductive and deductive approaches. Patterns and themes were explored \added{by discussion between both authors} throughout the coding process and after creating the code book. We considered the participants' responses to assess the fit between the designer intent and the taxonomy. Arising from the data, we found four types of affective goals that did not fit the taxonomy. Additionally, we studied the interviews to summarize how learning objectives would be helpful for participants. We anticipated that designers would find learning objectives difficult and we found several themes that gave us a deeper understanding of this. Finally, and most importantly, differing views arose on how some designers thought about affective intents.

% redacted
In the interest of protecting the anonymity of our participants, we have redacted and abstracted details from their interviews, replacing details with more general terms in [brackets] that still keep the essential meaning of the quote.

\subsection{Thematic Analysis}

% general overview of learning objectives
During the interview, we asked participants to create learning objectives for their visualization. All participants created at least one affective learning objective (max = 7, average = 3.7, median = 3). All but one also created a cognitive learning objective (max = 6, average = 2.9, median = 3). The fact that everyone reported an affective objective is unsurprising due to our recruitment strategy, but is certainly not representative of the population of communicative visualizations. \added{There was a wide range of learning objectives throughout the taxonomy, from a low-level ``\textit{the viewer will perceive the importance of the lockdowns}'' to a high-level ``\textit{the viewer will evaluate, question, and modify their own value-orientation system in comparison to the value-orientation system of democracy.}''}

% common topics / areas
There were some common themes across the topics of affective visualizations. Many of the visualizations were in the area of, or adjacent to, social justice or advocacy, covering topics that were political, controversial, or emotional. Given the sampling period, the most common topics were climate change and COVID-19. Other topics included refugees, racial justice, and policy changes. 

\subsubsection{Intent vs. Taxonomy}

% self-report
One of our main areas of inquiry was to see if the adapted framework of Bloom's affective learning objectives is a good fit for designers' intent. We asked our participants to self-report if they felt like the learning objectives fit their intent: \added{most} (9) participants responded yes, one responded \textit{no}, one responded that \textit{only the affective objectives fit}, and one responded that \textit{only the cognitive objectives fit}. 
We explore the \textbf{cases where the designers' intents are not translatable} to affective learning objectives to explore the limits of the taxonomy and to get a better idea of the entire affective space. 

% data art
One area where intent was hard to map to the taxonomy was in the case of \textbf{data art}. Some of our participants that identify their work this way indicated that the objectives didn't fit their intent. They described looking to provoke a reaction or evoke an emotion from their audience and were not necessarily interested in influencing their attitudes or values. One participant described this as, \textit{``it's only about an emotional response. You're trying to evoke something$\ldots$ regardless of what that emotion is.''} With a focus only on emotion, with no higher-level goal, we find that the taxonomy is not a good fit for solely `data art intents.' 

% personal project
The participant that responded that neither the cognitive nor the affective learning objectives fit her intent had created her data visualization as a \textbf{personal project}. Her intent was focused on goals for herself, not for sharing out to an audience. She described it as \textit{``when I make my own personal projects, it's more about me going through the process. It's for me.''} The participant wanted to learn about her own personal data and she created the piece as a way to push the limits of data visualization in an innovative way. Although the data visualization does evoke emotion and can influence the audience's appraisals, attitudes, and values, this was not the driving motivation for creating it. 

% as an object
A few of our participants had, amongst other affective goals, an intent for the visualization to be \textbf{an object in and of itself}. These visualizations were described as \textit{``a tribute''} or \textit{``a commemoration.''} %Similarly, a visualization can be described as ``a conversation starter''.
Conceptualizing the visualization as such an object may lead to affective goals---\textit{the viewer will consider [the tragedy]}---but the object itself did not directly represent a learning goal. 

% as a process
Finally, we found that act of creating the visualization was valuable \textbf{as a process}. One participant described his design process as \textit{``I try to express in some form or the other. So poetry is one form. Data visualization, I think, is similar.''} The act of data visualization can be an emotional process for the designer. Knitting a scarf of COVID deaths \cite{Schwab_2019} or temperature data \cite{Schwab_2019} can be a way to express grief and process emotions. Lupi described a goal of a collaborative project as: ``Through my work, I thought, I could try to help her process and communicate her emotions'' \cite{Lupi_2018}. In these cases, it's about the journey, not the destination. 

\subsubsection{Learning Objectives are Helpful}

% intro
In the semi-structured interview, we asked the question, \textit{``In what future situations would you use learning objectives?''} In response, we got a variety of answers about potential ways learning objectives would be useful for creating data visualizations. In general, \added{most participants (9)} indicated that learning objectives would be helpful in some way. The responses focused on two things: what the learning objectives would be helpful for and when in the design process they would be helpful. 

% for What?
\textbf{Helpful for what? }
% Novel / big projects
We found that some participants would only find learning objectives helpful for \textbf{big or novel projects}, such as \textit{``larger projects''} that they want to \textit{``show in a different way how it was never shown before''} or projects that are \textit{``more time consuming that$\ldots$ was a result of many weeks, a few months or so. It's more special to do something different than bar charts.''} They commented that learning objectives wouldn't be helpful for a simple bar chart that is straightforward and commonly used. 
% Communicating
About half \deleted{(7) }of our participants create data visualizations in a collaborative setting, involving communication amongst multiple people. Some of our participants mentioned that learning objectives would be helpful for \textbf{communicating with clients or a team}. One participant described them being useful \textit{``to ask my colleagues about, okay, well, what do we want the readers to get?}'' When working with a client, learning objectives could be integrated into a client brief. The purpose of a client brief is to outline the specifications, scope, and goals for the project. These are important to create clear expectations for client wants/needs and designer deliverables. One participant said,\textit{ ``I could 100\% see this integrated in a creative brief.''} 
% For design process?

% for When?
\textbf{Helpful when?}
Participants indicated that learning objectives would be helpful throughout the entire design process.
% Initial planning stages
The most common response was that participants were likely to use learning objectives in their \textbf{initial planning stages}. Participants indicated that creating learning objectives would be helpful right at the beginning of the design process. Participants mentioned, \textit{``it's very good to have these learning goals upfront''}, \textit{``I think it's very useful, right at the beginning of every project}'', and \textit{``I think these are highly useful in planning something new.''} In defining learning objectives at the beginning, they could refer to their goals while they are creating the visualization. Learning objectives would provide clarity, and turn a messy design process into clear goals.
% Choosing amongst designs
Some participants mentioned that learning objectives would help with \textbf{choosing amongst visualization designs}. This observation complements our past work showing that learning objectives can help designers choose a more effective design \cite{LeeRobbins_He_Adar_2022}. One participant indicated his design process as \textit{``random work trial and error''} and that \textit{``this will really make it easier, not as many trials before I reached the close. I think these are helpful.''} Although he mentioned that learning objectives are hard to create, he anticipated that they would be worth it because they would make the design process easier. 
% Evaluation
One participant noted that they could use learning objectives to \textbf{evaluate} their visualization once it's completed, \textit{``you can test against these goals, you can test if they, if they still hold up.''} However, many participants seemed to be \textbf{unsure of how to assess} if their visualization was successful in their affective goals. After the participant created learning objectives, we asked \textit{``Looking back at your visualization, do you think that it succeeds in these learning objectives?}'' Some participants didn't even know where to start with this, replying with \textit{``I'm not sure how to gauge that''}, \textit{``I don't know, that's something that I wouldn't be able to measure,''} and \textit{``Um, I think so. But um, yeah. I don't know, how could I be sure?''} Evaluation is an area where learning objectives could be helpful. Learning objectives could be translated into assessments, which could be used to evaluate if the desired response was achieved in the audience. We discuss a few options for assessments in Section \ref{discussion}.

\subsubsection{Learning Objectives are Difficult}

% intro/overall 
As we first noticed in our study of cognitive objectives, most participants found learning objectives difficult to create \cite{Adar_Lee_2020}. Anticipating this, we specifically asked our participants how hard it was to create learning objectives. \added{Almost all participants (11)}\deleted{Out of the 12 participants, 11} reported that they found some difficulty in creating the learning objectives. For most, this was the first time they had encountered the idea of learning objectives, let alone attempting to create them for their visualizations. Generally, participants reported that the affective domain was more difficult than the cognitive domain, with only one reporting the opposite.

% less aware of affective intents
One reason that the learning objectives, especially the affective ones, are hard to create, is that designers haven't given them much thought previously. Some participants indicated that they always do think about their intent, generally in other terms, or in the form of a client brief. However, others indicated that they are \textbf{not aware of their affective intent}. They mentioned having an affective intent, but they do not explicitly think about it. One participant describes this as \textit{``%I was more aware about the cognitive objectives$\ldots$ 
the affective learning objectives $\ldots$ is very interesting, because we use them, but$\ldots$ %we didn't really do it in a very, let's say. It's kind of in the subconscious. 
We don't speak about it consciously.''}
% frowned upon
Furthermore, even designers who are aware of their affective intent feel a stigma against explicitly labeling or talking about it.  
\begin{quote}
  \textit{``In journalism $\ldots$ it's sort of frowned upon to think about $\ldots$ or I guess to talk openly about the idea that$\ldots$ we're, like, trying to influence how people think or feel about a certain thing or $\ldots$ what actions we want them to take.''} 
\end{quote}

% verbs are difficult
Several people mentioned that some of \textbf{the verbs were difficult} to use. Participants commented, \textit{``it was quite hard to, like, use the verbs. Like, I really want to use the verbs, but I have trouble trying to form a sentence that fit to the ideas I have in my head.''} Because the taxonomy is based on educational goals, the verbs were often phrased in the positive direction---``the viewer will enjoy [the topic]''. These types of goals are more suited for the classroom, where the teacher wants the students to gain an appreciation or a passion for their subject. However, many of the affective visualizations that we've seen are about negative topics, such as climate change, deaths, and  injustices. This incongruence between verbs and topic made it harder for one of our participants to use the verbs. Instead of thinking of her objective as ``the viewer will accept [that we need to treat people better]'', it was more natural to frame it as ``the viewer will reject [the terrible treatment of people]''. The visualization story was about a tragedy, and the tone was negative and dark. The positive verbs were not a good match for her intent and the way she was designing the visualization. 

% ambiguity between affective VS cognitive
When creating learning objectives, there was an \textbf{ambiguity between the cognitive and affective domains}. After the explanation of learning objectives, some participants mentioned \textit{``I may$\ldots$ have difficulties to separate the cognitive ones from the affective ones.''} and \textit{``[t]hey go hand in hand, I would say I find it hard to separate them very, very easily.''} While creating cognitive learning objectives, goals arose that were better suited for the affective domain, and vice versa. 

\subsubsection{Differing Views on Affective Intent}

% intro
Previously, we discussed how the design choices that a designer makes can have an influence on how the audience will interpret and react to the message of the visualization. Throughout our interviews, participants had varying viewpoints on how they viewed the \textbf{responsibility of designer influence}. The viewpoints ranged from designers attributing full autonomy to the audience, the designers having an intent but acknowledging the variability within the audience, and designers taking full responsibility for the communication to the audience. Most participants fell in the middle of this range, but we also had some who were more towards one side than another. 

% no responsibility
Some designers disagreed with the idea that their visualization should affect their audience's values or behaviors. These designers talked about the audience as \textit{``it's everybody's own kind of decision to make up their mind''} or  \textit{``I'm not forcing anyone with my visualization that you should modify your [behavior]. I am not the person responsible for that.''} These designers thought about their role as only creating the visualization. After that, it's up to the viewer to engage with it. These designers felt like the affective learning objectives didn't fit their intent as well as the cognitive learning objectives did. One designer frames his intent as informing and \textit{``just show[ing] the facts''} clearly. Another designer frames his intent as \textit{``the viewer will consider [the phenomenon].''} %AL and JS
% even though it does influence
Even though these designers do not take responsibility for the audiences' outcome, they do agree that their visualization \textit{can and does} influence a person's attitudes. One designer noted that he received audience feedback indicating exactly that---that his visualization influenced their attitude. This designer agreed that his visualization does achieve the affective learning objectives, even though he disagrees that influencing their attitude was his intent. Another designer agrees that the ideal response that he would love to see in an audience would be for them to change their behavior. In fact, he mentioned specific design decisions that he made to increase the chance that the audience would change their behavior. However, he disagrees that changing the audience's behavior was his intent. 

% middle ground - audience agency, individual differences 
Most designers fell in the middle of the range, where they agreed that they had goals to influence their audience, but also acknowledged the individual differences within the audience would moderate each person's reaction to their visualization. These participants mentioned that not everyone is going to react in the same way, or that the audience has some agency in how they engage with the visualization. The participants said, \textit{``It's really more like, here's an information that I think is worth knowing and sharing, and you know, engage with it the way you'd like to engage with it''}, \textit{``Viewers will find this situation unacceptable, but that's really quite like more personal. I guess not everyone would find a situation unacceptable''}, and one compares his role to a radio host---\textit{``your audience is in control of the volume button, and they can put you on silent every time. So you have only the indirect control, you have to assume what your audience likes, and how you can convince them.''} 
% consequences
In a slightly different vein, some designers see a difference between outcomes that occur directly from information in their visualization and outcomes that occur beyond that. These outcomes might arise as a \textbf{consequence} of their visualization, but are not directly related. 
\begin{quote}
    \textit{``[T]he viewer will understand the [cognitive information]$\ldots$ As a consequence, they will [take an action]. But the project is not explaining why they have to [take an action]. [It's] just that is coming out of being exposed to the numbers and to, to kind of hopefully have the visualization having an impact in them. [It's] not a learning, [it's] a consequence, I would say.''}
\end{quote}

% far end
At the far end of the range, we had some participants describe the \textit{responsibility} that they have on influencing the audience. This viewpoint is that the designer is responsible for communicating the message. If the audience doesn't get the message, then the fault lies with the designer, not the audience. One participant described this as \textit{``How do you make sure they understand and you own that message failure? If that message fails, it's on you as the author.''} Another participant described this responsibility as not just providing information, but doing it in a way that is easily understood and not overwhelming. 
\begin{quote}
   \textit{``Because one of the responsibilities that we have, as visual journalists, is not just to like put all the information out there. It's also to direct the viewers eye, and direct the viewers attention$\ldots$ I take very seriously the responsibility like not to confuse people and not to overwhelm people.'' }
\end{quote}
These two participants did not specifically discuss affective intents in the context of this theme. They were either speaking more broadly about communication as a whole or at least more in the realm of cognitive efficiency. 
% conclusion
Most designers acknowledge that individual differences within the audience can cause a range of outcomes. Some use that as an excuse to absolve oneself from the responsibility of conveying a message. Others see it as their responsibility to communicate to everyone in the audience regardless of their background. 
\section{Revised Taxonomy}

% Introduction
In response to our interviews, we refined the original learning-focused affective taxonomy to better fit affective visualization objectives.

\begin{figure}[htbp!]
  \centering
\includegraphics[width=.9\linewidth]{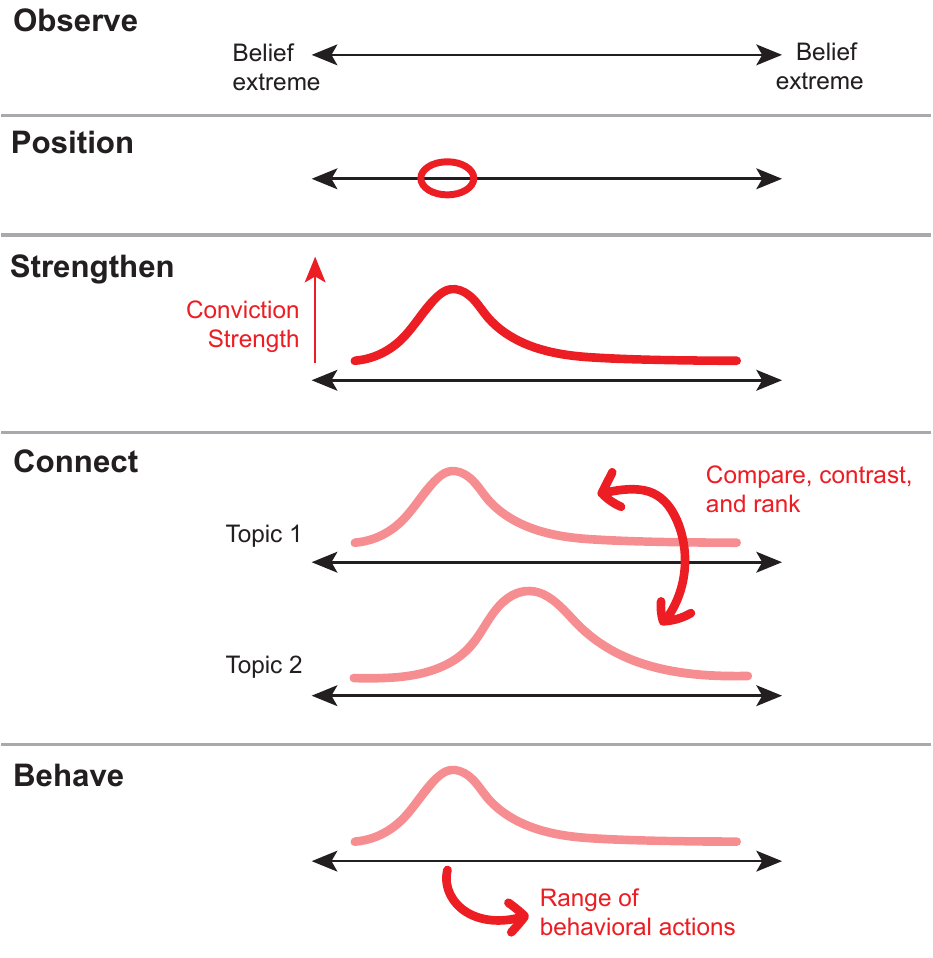}
  \caption{The revised affective verb taxonomy based on our interviews}
\label{fig:afftax}
\end{figure}

\subsection{Affective Verbs}

% Verbs
In our revised taxonomy, we have revised the verbs to be: \textit{observe}, \textit{position}, \textit{strengthen}, \textit{connect}, and \textit{behave}. The affective intents start at \textit{observe} the range of beliefs, then to \textit{position} oneself within that range of beliefs, then to \textit{strengthen} one's belief, then to \textit{connect} and compare several beliefs with each other, and finally to \textit{behave} consistent with that belief (see Figure~\ref{fig:afftax}).

In \textbf{\textit{observe}}, the intent is simply to make the viewer aware of the range of beliefs. At this beginning stage, the designer is not trying to influence the viewer to take one side or another. This is similar to \textit{perceive}, the lowest level of Bloom's taxonomy. When the viewer is simply observing the belief, there is minimal to no emotional response. For example, our goal may be to make the audience aware that there is a range of attitudes from ``pollution is the biggest health problem in the US'' versus ``pollution does not cause health problems.'' More realistically, this kind of objective may make a few `points' in the attitude space known to the viewer (e.g., a more restrained `point': ``pollution is a significant health problem in the US''). Additional verbs within this level would be \textit{perceive, see, consider, watch}, and \textit{attend to}.

The next level, \textbf{\textit{position}}, is when the designer intends to point out a specific place along the range of beliefs where they would like the viewer to be. At this stage, the viewer advances from observing the attitude to placing themselves onto the range of attitudes. This could be that the designer wants the viewer to self-reflect on their own prior attitude of where they land within the range. Alternatively, the designer could have a goal to position the viewer in a specific area of the range of attitudes. With `position,' there is no strong conviction, just a weakly held belief. An example of a learning objective could be ``the viewer will \textit{agree} that pollution is a major cause of negative health outcomes,'' placing them on one side of the attitude spectrum. Additional verbs within this level would be \textit{agree, accept, prefer, endorse,} and \textit{adopt}. 

The third stage, \textbf{\textit{strengthen}}, is when the designer tries to increase the strength of an attitude. Here, the designer doesn't want the audience to only agree with the belief, but to deeply support it. If the viewer already holds the specific belief, the goal would be to increase the intensity of the conviction. This level is similar to Bloom's level of \textit{believe}, where the student is not only accepting a value, but holding it more deeply and ingrained. As the strength of the conviction increases, so does the emotional response of the viewer. An example of a learning objective could be ``the viewer will \textit{trust} that pollution is a major cause of negative health outcomes.'' Additional verbs within this level would be \textit{intensify, support, believe, trust, internalize}, and \textit{commit to}. The distinction between \textit{position} and \textit{strengthen} is not a `clean' separation. Often, we will do both things nearly simultaneously (form an attitude and internalize it). However, the distinction allows us to acknowledge that at this level the intent of a visualization may be to \textit{weaken} a particular attitude and \textit{strengthen} another (i.e., shift the viewer from one position to another).

The fourth stage, \textbf{\textit{connect}}, is when the designer wants the audience to compare two of their beliefs. In this stage, the intent is for the viewer to compare, contrast, and rank multiple beliefs. The viewer is making judgements about which beliefs are more important to them, similar to Bloom's level of \textit{Organization}. An example of a learning objective could be ``the viewer will \textit{compare} their attitude toward sustainable eco-friendly options to their attitude toward convenient plastic options.'' Additional verbs within this level would be \textit{compare, contrast, rank, connect, sort, rate}, and \textit{evaluate}. 

The final stage, \textit{\textbf{behave}}, is when the designer wants the audience to act in ways consistent with the desired belief. Designers frequently have this goal, as demonstrated in ``calls to action'' often listed at the end of a visualization or story. Behaviors are the most visible indication of an affective outcome, yet these could be the most difficult to achieve. An example of a learning objective could be ``the viewer will \textit{buy} less plastic.'' Additional verbs within this level could be \textit{demonstrate, practice, act}, or a specific action verb related to the topic of the visualization. 

% Audience considerations
The audience's prior knowledge, beliefs, and backgrounds will naturally influence what type of verbs will be effective. When communicating to an audience that holds an opposite belief, the designer may be more successful with an \textit{observe} or \textit{position} goal, rather than having an outcome of a behavior. On the other hand, when communicating with an audience that already agrees with a belief, it will be more likely for a designer to strengthen that belief or inspire action for it. 

\subsection{Affective Nouns}

% Nouns
In the revised taxonomy, we redefined the lower levels of the nouns as \textit{appraisal} and \textit{attitude}, while maintaining the upper levels of \textit{value} and \textit{value system}. We found that the difference between opinion and attitude (and belief) were ambiguous and confusing to participants. There are many different accepted definitions for these terms in the literature, so we will more specifically describe how we are operationalizing the nouns for our revised taxonomy.

The lowest level of the noun taxonomy is an \textit{\textbf{appraisal}}. This is the level closest to the data and most straightforward to convey to the audience. An appraisal starts with a cognitive insight. For example: ``the United States is historically the largest contributor of carbon emissions.'' This is an unemotional fact as represented in the data. The lowest level of an affective response would be to associate the insight with a ``value judgement'' (i.e., that the data observed means something good or something bad). For example, an appraisal of this data could be ``the United States is historically the \textit{worst} contributor of carbon emissions.'' The change from \textit{largest} to \textit{worst} moves this simple fact to an affective evaluation. 

The next level would be \textit{\textbf{attitude}}, or a subjective disposition towards an object or phenomenon. An attitude is an inference that is one step away from the data, meaning that the data does not directly represent the attitude. This level is hierarchical, in that an attitude can be based on multiple appraisals, although there could be just one portrayed in the visualization. An example of this could be ``the United States is responsible to address climate change'' or ``carbon emissions are a major problem for the world.''

The upper two levels of the revised taxonomy are the same as the previous version: value and value system. \textit{\textbf{Values}} are a deeply held belief that are hard to change, are persistent over a long time, and are applicable across many situations. An example of a value could be ``restorative justice'' or ``equity.'' \textbf{\textit{Value systems}} are a set of beliefs held by a group of people. An example of a value system could be ``democratic'' or ``abolitionist''. There are various value system frameworks that may be used (e.g., Moral Foundation Theory~\cite{haidt2012righteous}).

% Audience
As with the verbs, the background and prior knowledge of the audience will impact what belief levels the designer would be likely to influence. A person's value system will moderate their acceptance of values, attitudes, and appraisals. If their value system is incongruent with the visualization, then the viewer may ignore, reject, or oppose the data. If the value system is in line with the visualization framework, then the viewer may be more willing to update their beliefs. 

\section{Affective Analysis: Juneteenth}

%Introduction
In this section, we conduct a critical analysis of an affective visualization to illustrate the use of the affective learning objectives taxonomy for a range of objectives. A critical analysis could be conducted with an external gaze. A critique of a visualization could cite design decisions and context clues for what learning objectives the visualization could potentially be pursuing, as well as where the visualization succeeds or fails. For this section, one of the designers of this visualization was a participant in our interview study. Thus, we have confirmation that the objectives stated were actually applicable. However, as the interview focused on framing intent in Bloom's taxonomy, we have translated the goals to the revised taxonomy, but have stayed faithful to the original intent. In our interview, we only talked to one of the authors, and the quotes below largely reflect her viewpoint on creating this visualization. We supplement this with quotes from a published conversation between the two authors that detail their process and goals for this visualization~\cite{Cogley_McRae_Blog_2020}. \added{Finally, we chose this example because of its range of learning objectives, but have included more examples at \url{visualobjectives.net}.}

\begin{figure}[ht]
  \centering
\includegraphics[width=.9\linewidth]{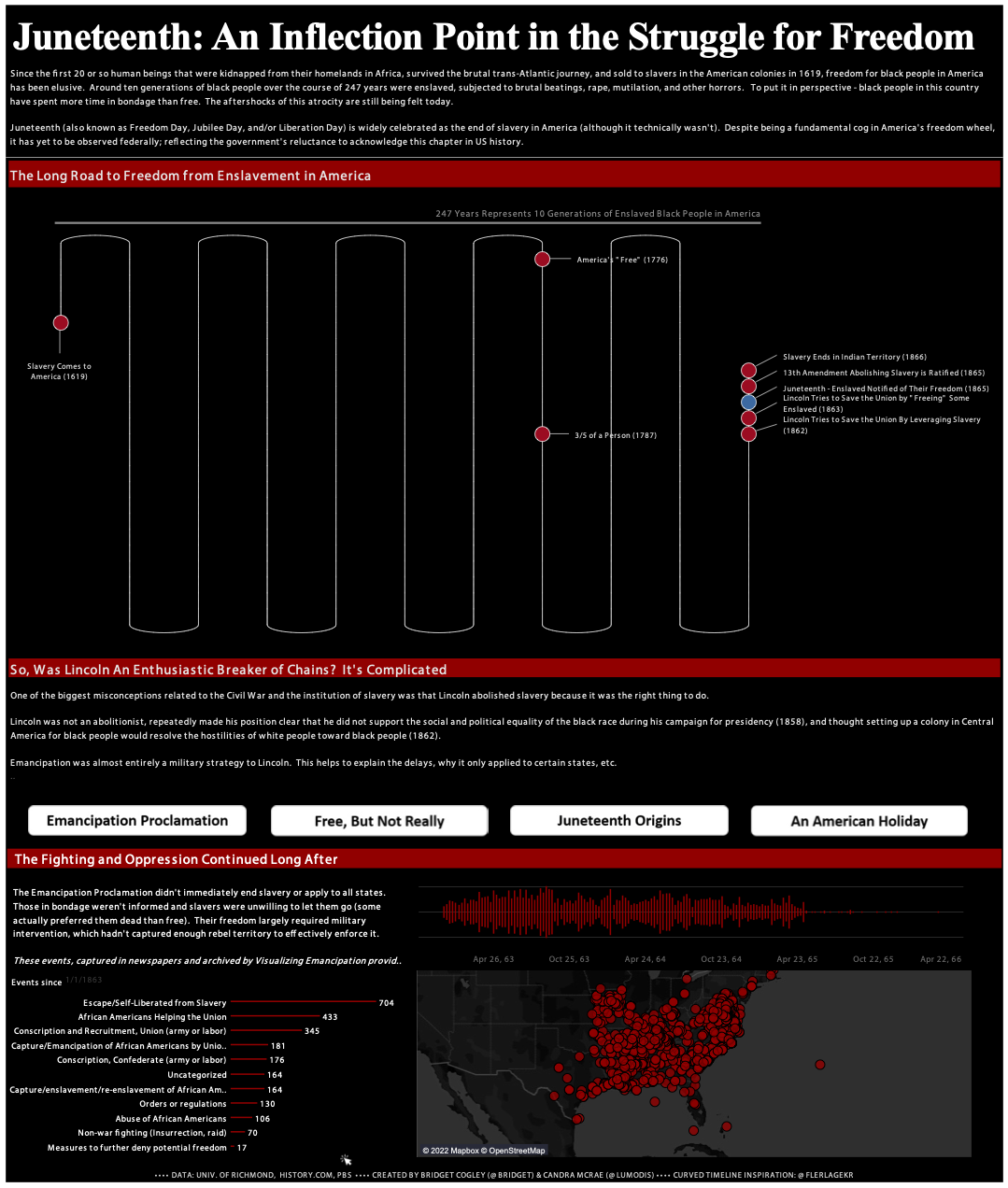}
  \caption{An affective data visualization showing the timeline to freedom for black people in America \cite{JuneteenthViz}. A larger version is included in our Supplementary Materials.}
\label{fig:juneteenth}
\end{figure}

% Description / Background
This visualization, titled ``Juneteenth: An Inflection Point in the Struggle for Freedom,'' communicates the realities \added{and misconceptions} of Juneteenth and the end of slavery in America in a visually provocative way (Fig. \ref{fig:juneteenth}). \added{During the American civil war, US President Abraham Lincoln issued the Emancipation Proclamation in 1863, outlawing slavery in the confederate states. But, the news didn't reach enslaved people in Texas until June 19, 1865, also known as Juneteenth.}
This visualization was created for affective intents. Our participant stated, ``\textit{it's not really a cognitive problem. It's an emotional problem}.'' %She created many learning objectives, both in the cognitive and affective domain. For the sake of space, we will only discuss some of them in this section. 
\deleted{This visualization was created in collaboration by two data visualization designers in June 2020, during the time when discussions of social justice were occurring, and before Juneteenth was officially designated as a federal holiday. }

% Observe
At a lower level, one of the goals of the visualization was to have the viewer \textit{observe} the unjust and slow delay of the timeline to freedom for black people. This was represented in the intentional design features of the curved timeline which starts at 1619 \added{(the start of slavery in America)} and ends at 1866 \added{(the end of slavery in America)}. Almost all of the timeline is empty, except for two events in the middle. One of the designers summarized the design choices and connected it to their intent, \textit{``The white space on the timeline was intentional because I wanted the reader to visually feel the time pass slowly between the onset of slavery and its abolition'' \cite{Cogley_McRae_Blog_2020}.} Additionally, the length of one of the turns on the timeline is representative of a generation (25 years). This was an intentional choice to make the connection of the years more relatable to the human scale. \deleted{This time period represents 10 generations in bondage.}

% Position
One of the goals of the visualization was, ``\textit{challenging of the mythology [of] $\ldots$ Abraham Lincoln, the Great Emancipator}.'' The mainstream understanding of the Emancipation Proclamation is that Lincoln freed the enslaved people for a moral and good cause. This visualization presents the reality of the situation---Lincoln issued the Emancipation Proclamation for financial, logistical, and strategic reasons for the civil war. In fact, \textit{``Lincoln was not an abolitionist, repeatedly made his position clear that he did not support the social and political equality of the black race''} \cite{JuneteenthViz}. For this visualization, the authors wanted to challenge the audience's prior beliefs (Lincoln had moral pursuits), and accept the reality they put forth (Lincoln used emancipation as a military strategy). Translated into a learning objective, this goal could be described as ``the viewer will \textit{accept} the narrative presented.''
The visualization had several elements to support this goal. The Tableau dashboard included text alongside the timeline to explain this narrative. The language accurately describes the reality of black people in slavery, using words like ``\textit{brutal}'', ``\textit{horrors}'', and ``\textit{atrocity}''. These words evoke a negative affective response, making the viewer painfully aware of the reality of slavery. Additional affective design elements include interactivity to unfurl each section, visual imagery and metaphor, and the overall dark tone of a black background with red colors.

% Connect
\deleted{This visualization opens up a conversation about broader values within oneself and valued in society within ones' cultural group. The author describes this as:
    \textit{``My cultural group typically puts the values here, do I agree? Do I have consensus with my cultural group? Or do I not and when I look at the broader sociological group, I realize that maybe my group is extreme.''} (quote from interview)
In terms of our revised taxonomy, we can structure this learning objective as, ``the viewer will \textit{connect} and \textit{compare} their own values to their value system of their cultural group.''}

There were several goals to move the needle slightly on values like equity and value systems as a whole. The viewer will (1) \textit{observe} or think about equity; (2) \textit{agree with} equity (in the long term) (3) \added{\textit{evaluate} their own value system and \textit{compare} it to their cultural group's value system (in the very long term).} The designer acknowledged that these goals, especially shifting values and value systems, are long term and not achievable with a single visualization. A quote from our interview highlights this line of thought:

\begin{quote}
    \textit{``And that may not happen right away with looking at this. But if we start to shift people here, they start to question the role of slavery, then they can start looking at other parts of this outside much later on, and really begin to think about it. And that's the whole goal of Juneteenth as a holiday, and that's the whole goal of this visualization is to progressively feed that value adjustment.''}
\end{quote}

% Behave 
Finally, the authors mention a call to action in their published conversation, inviting the audience to ``\textit{contact your Congressional representatives and sign the petition to make Juneteenth a federal holiday}'' \cite{Cogley_McRae_Blog_2020}. This call to action was not in the visualization. However, phrased as an affective learning objective, this goal could be framed as ``\textit{the viewer will take action in support of Juneteenth}.''

% collaboration
As mentioned previously, this visualization was a collaboration between two designers. Our participant mentioned that the collaboration consisted of conversations, dialogues, and an online whiteboard. Learning objectives could add to this discussion as a social object that the designers can create together, talk about, and refer to. Learning objectives might make it easier to collaborate with others by supporting a shared understanding of the goals of the project. 

\section{Discussion}
\label{discussion}

% stigma/controversy
In our interviews, we found that some designers disagreed that they had an intent to influence their audience. It's unclear if they were truly uninterested in affective intents or if they were uncomfortable with the idea of ``biasing'' their audience. In truth, this could vary from person to person. In our interviews, we saw that the designer's intent influences their design choices. Additionally, designers were aware that their visualization influenced (at least some of) the audience. Potentially, these designers perceive that an intent to persuade or influence is not a socially acceptable goal for data visualization. In the wider visualization community, there is a stigma or controversy that data visualizations should be ``objective.'' This perception of the community values might influence designers' perception of their own intent. However, we find that this value of ``objectivity'' is shifting within the community, with increasing discussions of subjectivity, bias, and affective intents.

We believe that it is important to acknowledge and model affective intents. In some sense, all visualizations are political and thus all have \textit{some} affective intent. It is problematic that people try to avoid their own subjectivity and deny affective intents when they have them. Accepting one's affective intents means that self-subjectivity is confronted. We see resistance to accepting subjectivity in data visualization because it is uncomfortable. Yet, there is value in accepting subjectivity. Once we acknowledge our own biases, we have the opportunity to self-reflect, be more aware of how they affect design, and lead us to create better visualizations. We agree that context is queen, meaning that it's not always favorable to include pathos techniques in visualizations \cite{DIgnazio_Klein_2020}. But recognizing affective intents within ourselves and within others' work will allow us to think more critically and deeply about affective intents and their use in data visualization. Building a language for affective objectives may better guide designers and critics. \added{We hope that designers will apply this framework to their own work and consider \textit{both} their cognitive and affective intents during their design process.}

% ambiguity between what is a technique/method and what is a goal? Emotion, humanizing
%Was in section 4, moved to discussion
There is some ambiguity of what is a technique in data visualization compared to what is a goal of data visualization. Our framework of affective learning objectives centers around reacting and responding to appraisals, attitudes, and values, and does not include a goal of simply evoking emotion. We consider \textit{evoking emotion} as a pathos technique used in data visualization in order to achieve a higher level goal, such as ``the viewer will strengthen a value''. For example, we considered \textit{humanizing the data} to be a technique in order to achieve a higher level goal, such as ``the viewer will support the refugees''. In the framework of rhetoric, an emotional appeal is part of making an argument. It does not make much sense for a designer to want the audience to feel sad, and then end there. There is an underlying reason that the designer wants the audience to feel an emotion. Maybe this is to consider the injustice of a situation, or to accept and internalize a value, or to act in response to the situation. In any case, the designer is not \textit{only} trying to make the audience sad. This is why we do not include emotion as a layer of the taxonomy. In fact, emotion could be thought of as a byproduct of a learning outcome. At lower levels of the taxonomy, there is not much of an emotional reaction (e.g., ``the viewer will consider the situation''). At higher levels of the taxonomy, there will be much more of an emotional response (e.g., ``the viewer will protest on behalf of human rights''). If an advocacy group is trying to get their audience fired up and emotionally charged, it's because they want them to take an action. If a journalist is trying to evoke emotion, it's because they want them to care about their topic.

\subsection{Future Directions}
% design process vs reflection
We explored the ways in which designers create learning objectives during an interview study. However, \added{this study only recruited designers \textit{after} they have created their visualizations. Therefore, we have explored learning objectives in a reflective context, by having designers think about a previous visualization they have already created.} There may be different considerations and insights that arise when designers use learning objectives during their actual design process. Future work should explore affective objectives as designers actively use them within their design process. We hypothesize that learning objectives would help designers create more effective affective visualizations. \added{Additionally, our interview study was focused on only the Bloom's version of affective learning objectives; we have not asked designers about their opinions on the wording of our revised taxonomy. Future work should specifically evaluate this version. }

% Data Art
While there is some overlap between affective visualizations and data art, we originally concluded that artistic goals are outside the scope of the taxonomy. In this paper, we mainly focused on affective \textit{intents}. However, the role of emotion in affective visualizations should be discussed more. The larger area of interest of how emotions are evoked in data visualization and what techniques in data visualization achieve that is an important field. A deeper investigation into data art may illuminate the connection between artistic endeavors, emotions, and reactions from an audience. We have thus far focused on \textit{intentional} objectives (hence our focus on design \textit{intents}). However, we believe it is worth considering when affective learning is achieved \textit{un-intentionally}.

% Materials to teach How to create Learning Objectives 
We found that creating learning objectives is difficult, especially affective ones. More materials should be produced to help designers traverse this area and clearly define their intent. Teachers have supporting materials for creating learning objectives, visualization design should also have supporting materials. An hour-long interview was a quick explanation, but more accessible materials should be created.

\added{Finally, we have not included ways to \textit{assess} whether, and to what extent, an affective visualization was successful in its goals. }
Part of the appeal of explicit definitions of learning objectives is they naturally lead to assessments. The main difficulties with evaluating affective intents is that they: shift on long-term timelines; are hard to accurately assess with typical multiple-choice questions; and outcomes can be behaviors. For assessments of \textit{position} goals, a yes/no response would suffice (e.g., Q: Do you agree that we should defund the police? A: yes/no). For assessments of \textit{strength} goals, a Likert scale could measure how strong the belief is (e.g., Q: To what extent do you agree that we should defund the police? A: strongly dis/agree, slightly dis/agree, neither agree nor disagree). For assessments of \textit{connect} or \textit{compare} values, ranking may be used (e.g., How much money would you allocate to each category given a fixed city budget?). For the final level, \textit{behaves}, a designer could ask for intentions to carry out behaviors (e.g., How likely are you to donate money to this cause?), measure indications of behaviors (e.g., track how many viewers clicked on a link to donate), or ask people later if they actually took action (e.g., Q: What actions have you taken in support of this cause this year? A: Donating; Signing petitions; Sharing materials; Contacting government representatives). 

A good measure will isolate the effect that the visualization has on the audience, rather than measuring the audience's prior beliefs. One way to do this is to ask pre- and post- questions. By contrasting a viewer's answer about their attitude before seeing the visualization and after, one could evaluate the success of the visualization. This method, and all multiple-choice methods, need to be designed carefully to ensure that they are measuring the intended phenomenon (effect of the visualization) and not something else (demand characteristics). Qualitative measures such as open ended responses, interviews, conversations, and observations can also give designers insight to how their visualization may influence their audience. Future work should expand on the idea of measuring and evaluating the success of affective visualizations.

\added{Learning objectives allow designers to explicitly identify their target goal. This can lead to thinking about what the result would look like if that goal was achieved. By considering their intent, designers can consider their audience's perception and reaction, even during design iterations. Even without the perfect assessment, getting feedback on designs in a casual way can give the designer valuable information how well they met their goal. }

\section{Conclusion}
If all visualizations are political, all visualizations are affective. To enable designers to design and talk about their affective goals, we develop a taxonomy for describing affective learning objectives for communicative visualizations. This work expands upon the idea that communicative visualizations are essentially a learning problem where the designer/viewer correspond to the teacher/student pairs. Our past work has identified ways of describing cognitive intents through a cognitive taxonomy. However, many visualizations--perhaps all--enable some affective learning. We begin with an existing taxonomy which we test against the intents of real designers. We update the taxonomy to better support visualization intents. Through a language of learning objectives, we offer a pathway for designers to think about their visualizations from both affective and cognitive lenses, contrast alternatives relative to specific articulation of their goals, and assess their work.

\acknowledgments{
We are grateful to the NSF for their support of this work through NSF IIS-1815760. We would also like to thank our interview participants for sharing their work and thoughts with us. Finally, we thank our reviewers for their helpful feedback.}


\balance{}

\bibliographystyle{abbrv-doi}

\bibliography{09_bib}
\end{document}